\documentstyle[12pt]{article}

\renewenvironment{thebibliography}[1]
{\normalsize
 \begin{list}{[\arabic{enumi}]}
 {\usecounter{enumi} \setlength{\parsep}{0pt}
  \setlength{\itemsep}{3pt} \settowidth{\labelwidth}{[#1]}
  \sloppy}}
{\end{list}}

\newcommand{\bea}{\begin{eqnarray}}
\newcommand{\eea}{\end{eqnarray}}
\newcommand{\beq}{\begin{equation}}
\newcommand{\eeq}{\end{equation}}

\def\msbar{\ifmmode{\overline{\rm MS}} \else{$\overline{\rm MS}$} \fi}
\def\drbar{\ifmmode{\overline{\rm DR}} \else{$\overline{\rm DR}$} \fi}

\begin{document}

\hfill\vbox{\baselineskip14pt
            \hbox{KEK-TH-371}
            \hbox{KEK Preprint 93-110}
            \hbox{UT-653}
            \hbox{August 1993}}
\vspace{10mm}

\baselineskip22pt
\begin{center}
\Large	Two-loop renormalization of gaugino masses in general
supersymmetric gauge models\footnote{Work supported in part by
Soryushi Shogakukai.}
\end{center}
\vspace{10mm}

\begin{center}
\large 	Youichi~Yamada
\end{center}
\vspace{0mm}

\begin{center}
\begin{tabular}{c}
{\it Theory Group, KEK, Tsukuba, Ibaraki 305, Japan}\\
\\
and\\
\\
{\it Department of Physics, University of Tokyo, Bunkyo-ku,
Tokyo 113, Japan}\\
\end{tabular}
\end{center}

\vspace{20mm}
\begin{center}
\large Abstract
\end{center}
\begin{center}
\begin{minipage}{13cm}
\baselineskip=22pt
\noindent
We calculate the two-loop renormalization group equations
for the running gaugino masses in general SUSY gauge models,
improving our previous result. We also study its consequence
to the unification of the gaugino masses in the SUSY SU(5) model.
The two-loop correction to the one-loop relation
$m_i(\mu)\propto\alpha_i(\mu)$ is found to be of the order of a few \%.
\end{minipage}
\end{center}
\vfill
\newpage

\baselineskip=22pt
\normalsize

It has recently been found \cite{gunif} that the experimentally
measured values
of three gauge couplings in the standard model are consistent with the
prediction of the supersymmetric (SUSY) SU(5) grand unified theory
(GUT) \cite{mm}:
the three running gauge coupling constants
$\alpha_i(\mu)=g_i(\mu)^2/(4\pi)(i=1,2,3)$
are unified in good precision,
\beq
\alpha_3(m_U)=\alpha_2(m_U)=\alpha_1(m_U)=\alpha_5(m_U) \label{eq1}
\eeq
at the GUT unification scale $m_U \sim 10^{16}$GeV.

In SUSY SU(5) model, there is another unification condition
which is related to
the SU(5) gauge symmetry. It is the unification of the
soft SUSY breaking running masses
of three gauge fermions (gauginos) $m_i(\mu)(i=1,2,3)$
at the same scale $m_U$. Namely, the relation
\beq
m_3(m_U)=m_2(m_U)=m_1(m_U)=m_5(m_U) \label{eq2}
\eeq
holds, apart from the threshold correction.

The one-loop renormalization group equations for the gaugino masses
take a very simple form \cite{1loopm}:
\beq
\frac{d}{dt}\left(\frac{m_i}{\alpha_i}\right)=0,\;\;\;\;\;\;\;\;
t\equiv \ln \mu ,\;\;\;\;\;\;\; i=1,2,3.  \label{eq3}
\eeq
By combining (\ref{eq2}) and (\ref{eq3}), we obtain the
follwoing well-known results in the leading order \cite{1loopm},
\beq
m_3(\mu)/\alpha_3(\mu)=m_2(\mu)/\alpha_2(\mu)=m_1(\mu)/\alpha_1(\mu)
=m_5(m_U)/\alpha_5(m_U). \label{eq4}
\eeq
Note that the relations (\ref{eq3}) hold in any SUSY gauge models.

In this paper, we study the two-loop correction
to the unification condition (\ref{eq4}).
This correction, together with the one-loop threshold corrections
at $m_U$ \cite{hisano} and at the weak scale, gives the complete
next-to-leading order correction to the identity (\ref{eq4}).
There are several reasons for studying the correction of
this order. First,
since the unification of the gauge coupling constants in the
SUSY GUT has already been studied in the
next-to-leading order \cite{gunif,gunif2},
consistent treatment of the gaugino mass unification also needs
the two-loop renormalization group equations for the running
gaugino masses. Second, the next generation of $pp$ and
$e^+e^-$ colliders are expected to be able to measure
the gaugino masses accurately enough \cite{jlc} to test the
unification condition in this order,
just as in the case of the gauge couplings at
present $e^+e^-$ colliders. Therefore, the next-to-leading
correction to the identity (\ref{eq4}) will become
important in the future study of the unification condition
of the gaugino masses and the test of the SUSY GUT's.

In the previous paper \cite{mine}, we have shown that the relation
(\ref{eq3}) is violated
in the two-loop order in the model which contains only the
vector supermultiplets.
In this paper, we extend our previous analysis of the two-loop
renormalization of gaugino masses to general SUSY gauge models,
by including the contributions of chiral supermultiplets, Yukawa
couplings and soft SUSY breaking trilinear
scalar couplings (A-terms).
We also show numerical estimates of the two-loop correction
to the gaugino mass unification in the SUSY SU(5) model.

Let us first fix our framework. We consider the SUSY gauge
model with a semi-simple gauge group $G=\prod_i G_i$,
where $G_i$'s are simple subgroups.
The model contains chiral supermultiplets $\Phi_a$ in
the representations
$R_a^{(i)}$ for the subgroup $G_i$.
The superpotential is
\beq
{\cal W}=\frac{1}{6}y^{abc}\Phi_a\Phi_b\Phi_c . \label{eq5}
\eeq
The soft-SUSY-breaking term in the lagrangian is
\beq
{\cal L}_{soft}=-\frac{1}{6}A^{abc}y^{abc}\phi_a\phi_b\phi_c
-\frac{m_i}{2}\lambda_i\lambda_i +{\rm h.c.}, \label{eq6}
\eeq
where $\phi_a$ and $\lambda_i$ denote the scalar component
of $\Phi_a$ and the gaugino of the group $G_i$, respectively.
The Yukawa couplings $y^{abc}$ and the A-terms $A^{abc}$
are defined to be symmetric with respect to the indices
$a$, $b$, $c$. We have omitted lower dimensional terms
in (\ref{eq5}) and (\ref{eq6})
since they are irrelevant for our study.

The two-loop renormalization group equations for the gauge
coupling constants $g_i$ \cite{2loopg,mv1}
are then expressed as
\bea
\frac{d}{dt}g_i &=&\frac{g_i^3}{(4\pi)^2}(-3C(G_i)+T_i(\Phi))
+\frac{g_i^5}{(4\pi)^4}2C(G_i)(-3C(G_i)+T_i(\Phi)) \nonumber \\
&&+\sum_{j} \frac{g_i^3 g_j^2}{(4\pi)^4}4T_i(\Phi)C_j(\Phi)
-\frac{g_i^3}{(4\pi)^4}y^{abc}y_{abc}\frac{C_i(\Phi_c)}{d(G_i)}.
\label{eq7}
\eea
Here we adopt the notations
\bea
&& C(G_i)\delta^{AB}= f^{ACD}_{(i)}f^{BCD}_{(i)}, \;\;\;
T_i(\Phi_a)\delta^{AB}= {\rm Tr}R^{(i)A}_aR^{(i)B}_a, \;\;\;
C_i(\Phi_a)I=R^{(i)A}_aR^{(i)A}_a,  \nonumber \\
&& T_i(\Phi)=\sum_aT_i(\Phi_a), \;\;\;
T_i(\Phi)C_j(\Phi)=\sum_aT_i(\Phi_a)C_j(\Phi_a), \;\;\;
y_{abc}=(y^{abc})^*, \label{eq8}
\eea
where $f^{ABC}_{(i)}$ denotes the structure constant of the
group $G_i$, and $d(G_i)$ is the dimension of $G_i$.

We obtain the two-loop renormalization group equations
for the running gaugino masses
$m_i$ in the \drbar scheme (dimensional reduction \cite{dr}
with modified minimal subtraction \cite{msbar}),
by evaluating the two-loop diagrams for the gaugino
propagators in the Wess-Zumino gauge: see Fig.1.
The results are
\bea
\frac{d}{dt}m_i&=&\frac{g_i^2}{(4\pi)^2}(-6C(G_i)+2T_i(\Phi))m_i
+\frac{g_i^4}{(4\pi)^4}8C(G_i)(-3C(G_i)+T_i(\Phi))m_i \nonumber \\
&&+\sum_{j}
\frac{g_i^2 g_j^2}{(4\pi)^4}8T_i(\Phi)C_j(\Phi)(m_i+m_j)
-\frac{2g_i^2}{(4\pi)^4}y^{abc}y_{abc}\frac{C_i(\Phi_c)}{d(G_i)}m_i
\nonumber \\
&&+\frac{2g_i^2}{(4\pi)^4}A^{abc}y^{abc}y_{abc}
\frac{C_i(\Phi_c)}{d(G_i)}. \label{eq9}
\eea
The term proportional to $C(G_i)^2$ is the contribution from the
diagrams with only the vector supermultiplets, which has been
found in ref.\cite{mine}.

Here we comment on the renormalization scheme dependence
of our results. The two-loop renormalization group equations
for the gaugino masses are dependent on
the renormalization scheme, whereas those for the gauge
coupling constants are independent. In this paper,
we adopt the \drbar scheme \cite{dr} since
this scheme respects supersymmetry while the usual \msbar
scheme \cite{ms} does not. Indeed, by using the formulae
in refs.\cite{mv1,mv2}, we find that the two-loop
renormalization group equations for the gauge vector
couplings and those for the gaugino couplings to the chiral
supermultiplets do not agree in the \msbar scheme.

Our results (\ref{eq9}) can be expressed as equations for the
ratios $m_i/\alpha_i$, by using the equation eq.(\ref{eq7}).
We find
\bea
\frac{d}{dt}\left( \frac{m_i}{\alpha_i}\right) &=&
\frac{g_i^2}{(4\pi)^3}4C(G_i)(-3C(G_i)+T_i(\Phi))m_i
+\sum_{j}
\frac{g_j^2}{(4\pi)^3}8T_i(\Phi)C_j(\Phi)m_j \nonumber \\
&&+\frac{2}{(4\pi)^3}A^{abc}y^{abc}y_{abc}
\frac{C_i(\Phi_c)}{d(G_i)}. \label{eq10}
\eea
The right hand side of (\ref{eq10}) contains only the
two-loop contributions, in accordance with the one-loop
identities (\ref{eq3}).
We can clearly see that the simple relations (\ref{eq3}) are
no more valid at the two-loop level in general SUSY gauge models.

In order to examine consequences of the two-loop corrections,
we consider the minimal SUSY standard model with two higgs doublets
and three generations of quarks and leptons.
The renormalization group equations (\ref{eq10}) for the ratios
$m_i/\alpha_i$ then take the form
\beq
\frac{d}{dt}\left( \frac{m_i}{\alpha_i}\right) =
\sum_{j=1}^3
b^{(m)}_{ij}\frac{g_j^2}{(4\pi)^3}m_j
+b^{(m)}_{i, top}\frac{y_t^2}{(4\pi)^3}A_t, \label{eq11}
\eeq
with
\beq
b^{(m)}_{ij}=\left( \begin{array}{ccc}
398/25 & 54/5 & 176/5 \\ 18/5 & 50 & 48 \\ 22/5 & 18 & 28
\end{array} \right), \;\;\;\;\;\;
b^{(m)}_{i,top}=\left( \begin{array}{c} 52/5 \\ 12 \\ 8
\end{array} \right) .\label{eq12}
\eeq
Here we have retained only the contributions from the top-higgs
Yukawa coupling $y_t$ with with the correcponding A-term $A_t$.

We estimate the two-loop correction to the relation (\ref{eq4})
numerically by integrating eq.(\ref{eq11}), while
approximating the right hand side by its one-loop solution.
By using the following inputs $\alpha_s(m_Z)=0.12$,
$\alpha(m_Z)^{-1}=128$, $m_{SUSY}\sim m_Z$ and the unification
conditions (\ref{eq1}) and (\ref{eq2}),
we obtain
\beq
\left( \begin{array}{c}
m_1/\alpha_1 \\ m_2/\alpha_2 \\ m_3/\alpha_3
\end{array} \right) (m_Z)= \frac{m_5}{\alpha_5}(m_U)
\left[ 1-\left(
\begin{array}{c} 0.043 \\ 0.068 \\ 0.036
\end{array} \right) - \left(
\begin{array}{c} 0.87 \\ 1 \\ 0.67
\end{array} \right)\Delta_t \right] , \label{eq13}
\eeq
where
\beq
\Delta_t=\frac{0.09y_t^2(m_U)}{1+11y_t^2(m_U)}
\left[ 1+0.4\frac{A_t(m_U)}{m_5(m_U)} \right]. \label{eq14}
\eeq
The main contribution to the above corrections comes from the
diagrams with internal gluino lines (see Fig. 1).
The contributions from the top-higgs interactions can be comparable
to the gauge interaction contributions only if
$|A_t(m_U)/m_5(m_U)|$ is greater than 10.

{}From eq.(\ref{eq13}),
we can clearly see that the deviations from the one-loop
identities (\ref{eq4}) are of the order of a few \%.
There is a possibility to detect these corrections
by future precision measurements.

In summary, we have obtained the two-loop renormalization group
equations for the gaugino masses in general SUSY gauge models
in the \drbar scheme. We have found that the one-loop proportionality
relations (\ref{eq3}) and (\ref{eq4}) between the gaugino masses
and the gauge coupling constants are violated at the two-loop order.
The two-loop corrections to the relation (\ref{eq4}) have been
evaluated numerically in the SUSY SU(5) model and they are found to
be detectable in future precision experiments.

\section*{\large Acknowledgements}

We would like to thank K. Hagiwara, H. Murayama and Y. Okada
for fruitful discussions. We also thank Soryushi Shogakukai for
financial support.

\section*{\large Note added}

After the calculation was completed, we received a preprint \cite{mava}
in which the renormalizaton group equations (\ref{eq9}) have been given.
Our results agree with theirs. They have also discussed the
renormalization scheme dependence in detail.


\def\PL #1 #2 #3 {Phys.~Lett. {\bf#1}, #2 (#3) }
\def\NP #1 #2 #3 {Nucl.~Phys. {\bf#1}, #2 (#3) }
\def\ZP #1 #2 #3 {Z.~Phys. {\bf#1}, #2 (#3) }
\def\PR #1 #2 #3 {Phys.~Rev. {\bf#1}, #2 (#3) }
\def\PP #1 #2 #3 {Phys.~Rep. {\bf#1}, #2 (#3) }
\def\PRL #1 #2 #3 {Phys.~Rev.~Lett. {\bf#1}, #2 (#3) }
\def\PTP #1 #2 #3 {Prog.~Theor.~Phys. {\bf#1}, #2 (#3) }
\def\ib #1 #2 #3 {{\it ibid.} {\bf#1}, #2 (#3) }
\def\etal {{\it et al}.}
\def\eg {{\it e.g}.}
\def\ie {{\it i.e}.}


\vspace{20mm}
\section*{Figure Captions}
\renewcommand{\labelenumi}{Fig.\arabic{enumi}}
\begin{enumerate}

\vspace{6mm}
\item
Two-loop diagrams which contribute to the gaugino mass renormalization.
The wavy line, solid line without an arrow, solid line with an arrow
and dashed line represent the gauge vector, gaugino, chiral fermion
and chiral scalar, respectively.
\end{enumerate}

\end{document}